\documentclass[12pt,a4paper,final]{iopart}

\usepackage{iopams}  
\usepackage{graphicx,todonotes,microtype,cite}
\usepackage[breaklinks=true,colorlinks=true,linkcolor=blue,urlcolor=blue,citecolor=blue]{hyperref}

\begin{document}

\title{Imaging cold atoms with shot-noise and diffraction limited holography}

\author{J. P. Sobol and Saijun Wu}
\address{Department of Physics, College of Science, Swansea University, Swansea, SA2 8PP, United Kingdom}
\ead{brynsob@yahoo.co.uk, saijun.wu@swansea.ac.uk}

\begin{abstract}
We theoretically develop and experimentally demonstrate a holographic method for imaging cold atoms at the diffraction and photon shot noise limits. Aided by a double point source reference field, a simple iterative algorithm robustly removes the twin image of an $^{87}$Rb cold atom sample during the image reconstruction. Shot-noise limited phase shift and absorption images are consistently retrieved at various probe detunings, and during the laser cooling process. We consistently resolve less than 2~mrad phase shift ($0.4\%$ attenuation) of the probe light, outperforming shot-noise limited phase-contrast (absorption) imaging by a factor of 4 or more if the same camera is used without pixel saturation. We discuss the possible extension of this work for precise phase imaging of dense atomic gases, and for off-resonant probing of multiple atoms in optical lattices. 
\end{abstract}

\pacs{67.85.-d, 42.40.-i, 37.10.Jk,  42.50.-p}
\vspace{2pc}
\noindent{\it Keywords}: cold atoms, laser cooling, holographic microscopy, twin image removal, shot noise limited sensitivity.
\submitto{\NJP}

\section{Introduction}

Since the achievement of Bose-Einstein condensation~\cite{BECpub},  many-body physics in textbooks has been reproduced with beautiful experiments in laser cooling labs. Owing to its controllability and precision, the field of ultra-cold atoms
holds unique promise both for answering important questions in condensed matter physics, and to generate new physics. Breakthroughs in cold atom research are often accompanied with improved imaging techniques~\cite{ImageBreakthrough}. Recently, in situ imaging~\cite{Bakr09,Sherson10} has been developed to probe the shortest length scales of the confined, sub-$\mu$K quantum gases and is referred to as quantum gas microscopy.

In situ imaging of high density atomic gases suffers from detrimental effects related to resonant interactions mediated by photons. For example, state-of-the-art in situ florescence detection~\cite{Bakr09,Sherson10,Nelson07}
cannot image an optical lattice site with more than one atom occupancy without losing the extra atoms in pairs. Similarly, absorption imaging is perturbed by resonant dipole interactions, and the atomic density information cannot be faithfully retrieved via the Beer-Lambert law~\cite{Chomaz12}. To mitigate the effect one may saturate the atomic transition with a strong probe, resulting in images with excessive photon shot noise~\cite{Yefsah11,Othersolution}. Off resonant imaging~\cite{phasecontrast, Higbie05, Sanner11, Schley13, Kadlecek01, Turner04, Turner, TurnerThesis,Light} provides a solution to the density dependent line-broadening problem, since the variation of probe light phase shift due to line shape changes vanishes at large detuning. However, the magnitude of the phase shift due to a single atom reduces with detuning, while the phase shift sensitivity is limited by the photon shot noise as $\delta \phi \sim 1/\sqrt{N_p}$, where $N_p$ is the average number of probe photons detected by a pixel. Imaging with optimal use of the sensor depth and at the photon shot noise limit is essential for detecting the weak probe absorption/phase shift from few or single atoms. Since $N_{p}<N_{\rm max}$, the maximum pixel count, it is difficult with phase-contrast imaging to achieve a sensitivity below 5~mrad using standard cameras, and the amount of detuning allowed for off-resonant detection of cold atoms is correspondingly limited.

This work is motivated by an advantage of holographic microscopy~\cite{Gabor48}, that has to our knowledge been overlooked: It has the capability of detecting objects with weak phase shift/absorption (e.g. mrad/$0.1\%$ level), such as with off-resonant imaging of single atoms, where the probe photon shot noise may prohibit detection using standard imaging techniques with a regular camera (e.g. with $N_{\rm max} < 10^6$). Holographic microscopy~\cite{Gabor48} reconstructs a complex wavefront $E_s$ from a hologram $H$ that contains the interference pattern between $E_s$ and a known spherical wavefront $E_r$ (Fig.~\ref{fig1}A). It is well-known that holographic imaging can be free from lens aberrations even at large numerical aperture (NA)~\cite{Kanka11}, which can in principle be useful for imaging cold atoms at long working distances~\cite{ImageBreakthrough,Bakr09,Sherson10,Nelson07}. However, it is known that an image reconstructed in an inline holographic geometry (Fig.~\ref{fig1}A) is contaminated by an out-of-focus twin image~\cite{Gabor48}, restricting the method when imaging large objects such as a typical cold atomic sample.  In addition, to address the narrow atomic transitions a probe laser with long coherence length is required, leading to speckle noise in the hologram. To date, approaches implemented to overcome the twin-image problem in cold atom imaging fall into two categories: Firstly, spatial heterodyning can be explored to uniquely determine the phase of $E_s$ in an interferometric setup~\cite{Kadlecek01}. Indeed, outside of cold atom research, spatial and/or temporal heterodyning methods have been developed for shot-noise limited holography~\cite{Gross}. However, the heterodyning setup is complex and typically has a limited NA and spatial resolution. Secondly, with simple setups, defocus-contrast imaging~\cite{Turner04} and diffraction contrast imaging~\cite{Turner,TurnerThesis} exploit the monomorphous responsivity of atoms. However, these methods are limited to detecting an atomic sample with a uniform refractive index while its real part must be negative, and the methods do not completely suppress the so-called DC noise, as will be detailed in this paper. Proposals for cold atom imaging~\cite{Huang14,Ku11} with phase shifting (temporal heterodyning) holography~\cite{Yamaguchi97} promise twin image removal with fast reconstruction, but demand multiple and typically slow exposures.

In this work, we present the first demonstration of diffraction limited holographic imaging of cold atoms with photon shot-noise limited sensitivity.  We introduce a hybrid geometry (Fig.~\ref{fig1}B) to solve the twin image problem, and demonstrate shot noise limited holographic imaging in the presence of laser speckle noise. Furthermore, by taking advantage of point source holographic recording, we achieve $\sim 2$~mrad phase shift and $\sim0.4\%$ absorption sensitivities, beyond the photon shot noise limit imposed by $N_{\rm max}$ in standard imaging~\cite{foot:Superpix}. Our atom detection sensitivity is near the single atom level with resonant probing, which is achieved at NA=0.075 and thus a moderate resolution of $x_{\rm res}=5.2~\mu$m. By improving the spatial resolution to the wavelength limit, which would not require precision high-NA optics~\cite{Kanka11}, we expect dramatic improvement of the detection sensitivity, allowing precise phase imaging of confined atoms with single atom resolution.  With these features, we discuss the method for off-resonant imaging of high density ultra-cold gases, and for non-destructive probing of optical lattices with multiple atom occupancy per site.

Our approach to retrieval of $E_s$ from holograms is an extension of the Gerchberg-Saxton iterative method~\cite{review1} for localized samples with concrete supports~\cite{koren1991}.  As detailed here, by increasing the information contained in the reference field $E_r$, our hybrid geometry helps to improve both the speed and robustness of the $E_s$ convergence, even for samples with large spatial extent, signal near the photon shot noise level, and no presumption on the sample polarizability~\cite{Latychevskaia,Rong,Turner04,Turner}. The major disadvantages of our method include a factor-of-two reduction in signal compared to standard absorption/phase contrast imaging, a low dynamic range for atom signal detection which arises from speckle noise, and iterative reconstruction preventing instantaneous imaging. The first disadvantage is inherent to holographic detection techniques, while the second is associated with the narrow-line laser for imaging cold atoms. We discuss methods under investigation that have shown potential to overcome the 2nd and 3rd disadvantages.


\section{Hybrid holographic microscopy}
\begin{figure}[htb!]
 \centering
  \includegraphics [width=5in] {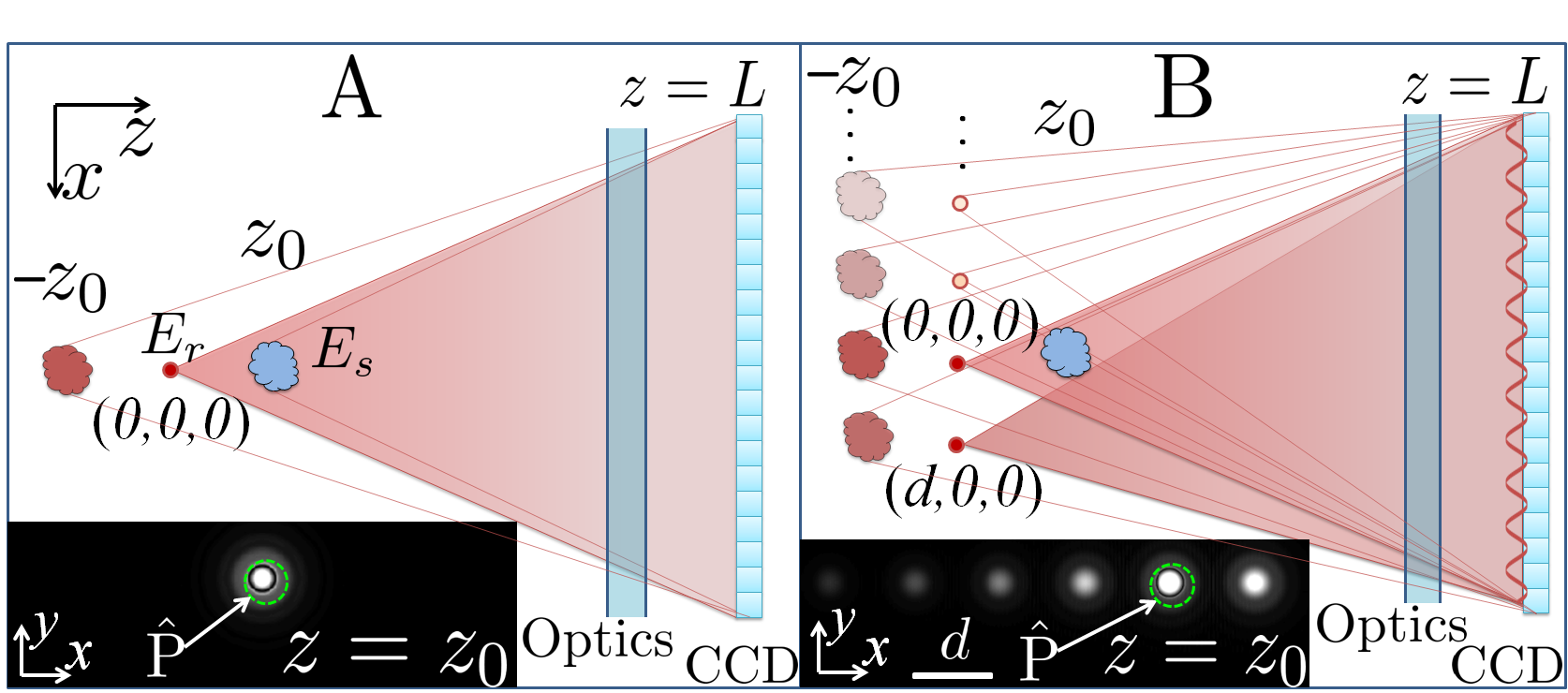}
  \caption{Schematics of a holographic microscope in the inline (A) and hybrid (B) geometries. A CCD camera records the hologram $H=|E_s+E_r|^2$. A one-step numerical reconstruction of $E_s$ leads to an image at $z=z_0$, and the twin images (at $z\approx-z_0$) and DC images (small disks at $z=0$) respectively. The insets display slices of $|E_s^{(0)}|$ at $z=z_0$ reconstructed from a simulated hologram. An aperture $\hat {\rm P}$ is used for iterative removal of the twin and DC images.
	\label{fig1}}
\end{figure}

Our microscope is depicted in Fig.~\ref{fig1}B, together with the traditional inline setup~\cite{Gabor48} in Fig.~\ref{fig1}A. To generate the reference wave $E_r=E_{r,1}+E_{r,2}$ in the hybrid setup, we add a second, ``off-axis'' point source $E_{r,2}$ at ${\bf r}_2=(d,0,0)$ displaced from the inline source
$E_{r,1}$ at ${\bf r}_1=(0,0,0)$ with light power $P_2=\eta P_1$, resulting in an interference pattern at the camera plane $z=L$. In both setups, the known wavefront $E_r$ interferes with $E_s$, the elastically scattered light from the sample located at $z=z_0$. The  intensity is recorded as the hologram $H=|E_r+E_s|^2$. With $H_0$ the hologram taken without the sample, an approximation of the 2D wavefront $E_s(L)$~\cite{foot:notation} is written as
\begin{equation}
\label{eq:htwo}
E_{H}=\frac{H - H_0}{{E^{*}_{r}}} = E_s + \frac{E_r E^{*}_{s}}{{E^{*}_{r}}} +
\frac{|E_s|^2}{{E^{*}_{r}}}.\label{Eq:EH}
\end{equation}

Using the angular spectrum method~\cite{Goodman} we propagate $E_{H}$ from $z=L$ to $z=z_0$, with a numerical propagator $\hat {\rm U} (z-L)$ that generally relates two wavefronts $E(z)$ and $E(L)$ through $E(z)=\hat {\rm U} (z-L) E(L)$. Here  $\hat
{\rm U} (z)=\hat {\rm F}^{-1}e^{i z  \sqrt{k^2-k_x^2-k_y^2}}\hat {\rm F}$, $k=2\pi/\lambda$ and $\hat {\rm F}$ is the 2D Fourier transform with $E(k_x,k_y,z)=\hat {\rm F} E(x,y,z)$. In addition to $E_s$ focusing at $z=z_0$, the 2nd and 3rd terms on the right hand side of Eq.~(\ref{eq:htwo}) focus at $z\approx-z_0$ and $z=0$, and are commonly referred to as the twin and DC images respectively. Optics between the sample and camera can be modeled for aberration correction. Without aberrations, the resolution of $E_s(z_0)$ is diffraction limited to $x_{\rm res}=\lambda \sqrt{w^2/4+(L-z_0)^2}/w$, with $w$ the camera width.

\begin{figure}[htb!]
 \centering
  \includegraphics [width=5in] {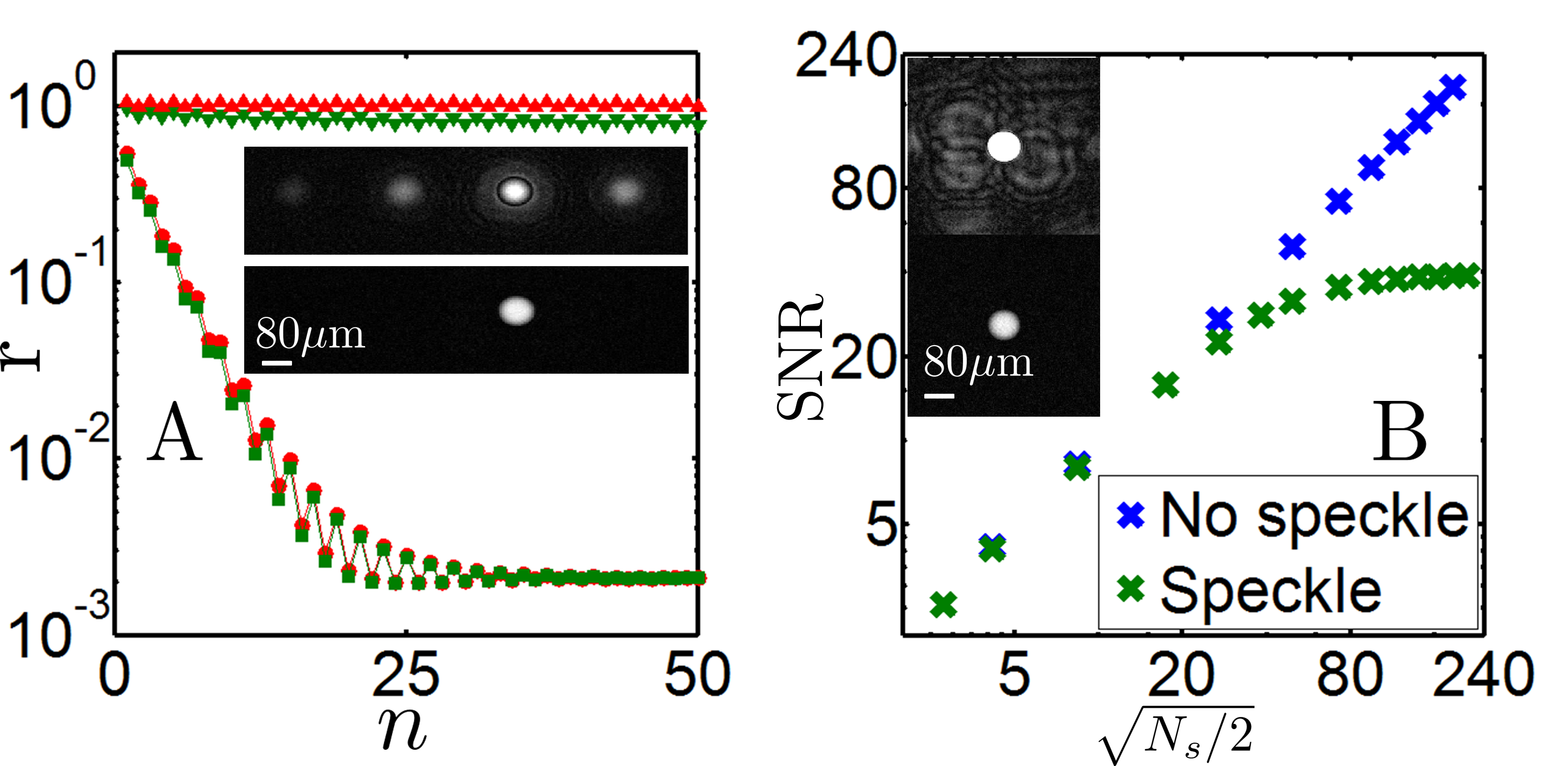}
  \caption{A) Convergence of the twin image removal algorithm for a simulated phase object. The residual $r$ is plotted vs $n$ for ${\rm F}=3.8$ and ${\rm F}=28$ with squares and circles respectively in the hybrid geometry ($\eta=0.16$), and with inverted triangles and triangles respectively in the inline geometry. Inset): The one-step $|E_s^{(0)}|$ (top) and converged $|E_s^{(n=50)}|$ (bottom) for ${\rm F}=28$, $\eta=0.16$. B) Simulated SNR of the converged $|E_s^{(n)}|$ vs $\sqrt{N_s/2}$ in the presence of photon shot noise (blue crosses), and of both speckle and shot noise (green crosses). Inset): the converged $|E_s^{(n=50)}|$ for a strong (top) and a weak (bottom) phase object in the presence of the same speckle noise.
 \label{fig2}}
\end{figure}

To isolate the real image $E_s(z_0)$, we notice that $E_s(z_0)$ is in focus at the sample location, while the twin and DC terms are spread out (Fig.~\ref{fig1}). With an aperture operator $\hat {\rm P}$ that sets the data outside of the sample area with a characteristic diameter $a$ to zero, the energy of the out-of-focus images can be removed iteratively with an algorithm similar to ref.~\cite{koren1991},
\begin{equation}
\begin{array}{l}
E_{s}^{(0)} = \hat {\rm U}_0 E_{H},\\
E_{s}^{(n+1)} = \hat {\rm U}_0 \hat {\rm C} \hat {\rm U}_0^{-1} \left( E_{s}^{(0)} -
\hat {\rm P} E_{s}^{(n)} \right).
\end{array}
\label{eq:Algorithm}
\end{equation}
Here $\hat {\rm U}_0\equiv \hat {\rm U}(z_0-L)$, so $\hat {\rm U}_0$ and $\hat {\rm U}_0^{-1}$ transform the wavefronts between the camera and sample planes. $\hat {\rm C}$ is a conjugation operator $\hat {\rm C} E=E^{*}E_r/E^*_{r}$, that converts any wavefront at the camera plane into that of its twin. Eq.~(\ref{eq:Algorithm}) assumes $|E_s|^2\ll|E_r|^2$ everywhere on the camera. When this is not valid, Eq.~(\ref{eq:Algorithm}) is modified to Eq.~(\ref{eq:Algorithm2}) which removes the DC term in Eq.~(\ref{eq:htwo}). With $E_{s}^{(n)}=E_s(z_0)+\delta E_s^{(n)}$ in Eq.~(\ref{eq:Algorithm}), the error $\delta E_s^{(n)}$ is easily shown to converge near zero (Fig.~\ref{fig2}A). Specifically, the residual energy fraction $r=\int
|E_s^{(n)}-E_s(z_0)|^2 d x d y/\int |E_s(z_0)|^2 d x d y$ decays with a characteristic constant $N_0\approx -1/\log(\epsilon)$,
\begin{equation}
\epsilon=\frac{\int |\hat {\rm P} \hat {\rm U}_0 \hat {\rm C}\hat {\rm U}_0
^{-1}\hat {\rm P} {\rm I} |^2 d x d y}{\int |\hat {\rm U}_0 \hat {\rm C}\hat {\rm U}_0
^{-1}\hat {\rm P} {\rm I} |^2 d x d y}.
\label{eq:exponent}
\end{equation}
Here $\epsilon$ is the fraction of energy of the out-of-focus twin image $\hat {\rm U}_0 \hat {\rm C}\hat {\rm U}_0 ^{-1}\hat {\rm P} {\rm I}$, that is contained within the aperture $\hat {\rm P}$ at $z_0$ (${\rm I}(x,y)=1$ is a 2D uniform wavefront so $\hat {\rm P} {\rm I}$ is the image of the aperture itself.). The aperture $\hat {\rm P}$ must be larger than the sample, thus $\epsilon$ can characterize the overlap
between the real and twin images.

To improve the convergence speed one needs to reduce the overlap. For the inline geometry (Fig.~\ref{fig1}A) a large sample needs an aperture with a Fresnel number ${\rm F}=a^2/2 z_0\lambda>1$, leading to a reduced diffraction effect and thus $\epsilon\approx 1$. In the hybrid geometry, the twin image is split into multiple copies displaced by ${\bf r}_2-{\bf r}_1$ (Fig.~\ref{fig1}B), an effect that can be understood by considering the corresponding term in  Eq.~(\ref{eq:htwo}) (second term on the right) with $E_r$ being a wavefront from a two-point source with power ratio $\eta$. It is easy to show that the twin image that is inline with the sample takes a fractional energy of $(1-\eta)^2$ for $\eta<1$.  Thus even for ${\rm F} \gg 1$, $\epsilon \approx (1-\eta)^2$ can still be small. In Fig.~\ref{fig2}A we plot the residual $r$ vs iteration number for a simulated phase object. For the hybrid scheme with $\eta=0.16$, the series converges quickly with $N_0=2.8\approx -1/\log{(1-\eta)^2}$ for both ${\rm F}=3.8, 28$. The non-zero final residual in the simulation is mainly due to the boundary artifacts in the FFT. In contrast, the inline scheme fails to converge within $10^3$ iterations even for ${\rm F}=3.8$.

So far, we have discussed reconstructing $E_s$ from noiseless holograms with a perfect $E_r$.  Major sources of noise are imperfect subtraction  $H-H_0$, aberrations and speckle noise in $E_r$ and $E_s$, and photon shot noise in $H$ and $H_0$. The subtraction is critical for reaching the shot noise limit, for which we have developed an optimization algorithm as detailed in section~\ref{secDetail}. Photon shot noise in $H$ and $H_0$ is proportional to $\sqrt{N_p}$ at each pixel with area $A_p$. Assuming an equal shot noise level in $H$ and $H_0$, the root mean square (rms)  $E_s^{(n)}$ shot noise level is found to be $\sqrt{2\hbar \omega /A_r Q \tau}$ ($\hbar \omega$, $\tau$, $Q$ and $A_r=x_{\rm res}^2$ are the photon energy, exposure time,  quantum efficiency, and resolution area respectively.), i.e. a  2-photon equivalent light field amplitude. Also as detailed in section~\ref{secDetail}, we found that the noise penalty to the twin image removal is small except when both $a/z_0\gg$~NA  and $\eta\ll 1$, a regime easily avoided in the hybrid geometry. The signal-to-noise ratio (SNR) of the converged $|E_s^{(n)}|$ is thus shot-noise limited to $\sqrt{N_s/2}$ with $N_s=\langle |E_s(z_0)|^2 \rangle A_r Q \tau/\hbar \omega$~\cite{Gross}. Similar to the limit in standard absorption or phase contrast imaging~\cite{lye1}, this SNR limit is decided only by the number of photons elastically scattered by the atoms, and is not sensitive to the probe detuning. While remarkably, the SNR is not sensitive to a change in $L$ in the holographic setup, and is only a factor of two less than the ${\rm SNR}\rightarrow \sqrt{2 N_s}$~\cite{lye1,Kadlecek01} in standard imaging where effectively $L\approx z_0$ within the depth of view. By taking holograms at $L\gg z_0$ to reduce the intensity of the point-source reference fields at the recording plane, the phase shift sensitivity, defined at SNR=1 within $A_r$, is improved from $\delta \phi=\sqrt{A_p/2 A_r N_{\rm max}}$ in standard imaging to $\delta \phi \approx\sqrt{2 \kappa A_p/A_r N_{\rm max}}$, with $\kappa=|E_r(L)/E_r(z_0)|^2$ (see section~\ref{secDetail}).

Addressing atomic transitions requires a laser with a long coherence length. Thus holograms of cold atoms can have significant speckle noise (Fig.~\ref{fig3}A), which is typically due to distant point-like scatterers. The speckle noise $E_{\rm speck}$ compromises our knowledge of $E_r=E_{r,1}+E_{r,2}+E_{\rm speck}$. Ignorance of $E_{\rm speck}$ leads to multiple copies of $E_s(z_0)$ shifted by the corresponding distances between the distant scatterers and ${\bf r}_{1,2}$, that blur the reconstructed $E_s(z_0)$ image (Fig.~\ref{fig2}B). In Fig.~\ref{fig2}B we plot the SNR of the converged $|E_s^{(n)}|$ vs $\sqrt{N_s/2}$, simulated in the presence of both speckle and photon shot noise. As expected, the reconstruction can be shot-noise limited if the speckle noise induced blurring of $E_s(z_0)$ is weak, as in the case of a small atomic sample~\cite{futurework}.
\section{Experimental Setup}

We demonstrate the hybrid microscope with a simple experiment: A $^{87}$Rb magneto-optical trap (MOT)~\cite{metcalf} is formed at the center of a vacuum glass cell. The atoms are cooled to $\sim 80~\mu$K and holographically imaged by a camera (pco.pixelfly usb with 1392$\times$1040 pixels, $x_{\rm cam}=6.45~\mu$m pixel pitch, quantum efficiency Q=0.1 at 780~nm, 14 bit dynamic range with a typical readout noise of 6 to 8 counts) placed outside the cell. To achieve sub-pixel resolution, we simply split each pixel in the camera-recorded hologram into $M^2$ new pixels, each with pixel size $\tilde{x}_{\rm cam}= x_{\rm cam}/M\leq x_{\rm res}$.

The probe, detuned by $\Delta$ from the D2 hyperfine $F = 2$ -- $F'= 3$ cooling transition ($\lambda$=780.2~nm), is derived from the output of an external cavity diode laser offset-locked to the D2 transition with $< 1$~MHz linewidth. The probe is frequency controlled by an acousto optic modulator before coupling into a polarization maintaining fiber. The output of the fiber is collimated and then split into two beams, which are focused by an aspheric lens into diffraction limited spots of 3.5~$\mu$m and 5.2~$\mu$m, resulting in the co-propagating wavefronts $E_{r,1}$ and $E_{r,2}$. A minimal amount of optics are mounted on a rigid platform to independently control the direction and diameters of $E_{r,1}$ and $E_{r,2}$. With the focal points ${\bf r}_{1,2}$ displaced by a measured distance of $d$=387~$\mu$m, an interference pattern with fringe periodicity 96~$\mu$m is captured by the camera at $L$=47~mm. With $z_0=3$~mm, we adjust the atoms location so that the atoms are only interrogated by the inline source $E_{r,1}$. To control the atom sample size, we vary the loading time of the MOT and ramp the magnetic field gradient from 10~G/cm up to 60~G/cm before imaging. The cooling laser detuning is at $-2~\Gamma$ ($\Gamma =2\pi\times 6.1$~MHz is the linewidth of the D2 transition). To test the spatial resolution,  a 1D lattice 32~$\Gamma$ detuned is pulsed to write a 20~$\mu$m period atomic density grating along $y$.  We image during and after cooling, at various $\Delta$, $\tau$, and probe intensities $I$.

\section{Results and discussions}

\begin{figure}[htb!]
 \centering
  \includegraphics [width=6in] {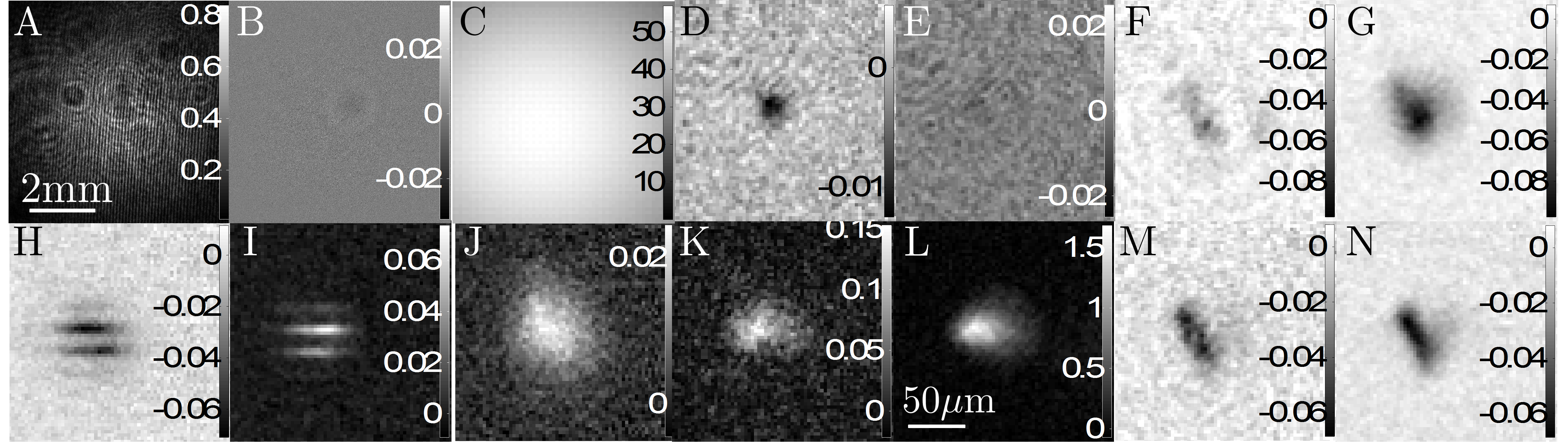}
  \caption{Holographic reconstructions (D-N), and intensity images (W/m$^2$) during the reconstruction process (A-C) . The hologram $H$ in A) and the  subtracted hologram $H-H_0$ in B) obey the $2$ mm scale bar. The $|E_{r,1}(z_0)|^2$ in C), and images in D)-N) obey the 50 $\mu$m scale bar. D) and E) [$\Delta = 30~\Gamma$, $\tau = 100~\mu$s]: Phase shift $\phi$ and absorption coefficient $\alpha$ images reconstructed from B). F) and G) [$\Delta = 1.9~\Gamma$, $\tau = 100~\mu$s]: $\alpha$ and $\phi$ images. 	H) [$\Delta = 1.9~\Gamma$, $\tau=20~\mu$s] and I) [$\Delta = -7.1~\Gamma$, $\tau=20~\mu$s]: $\phi$ images with $20~\mu$m fringes. J) [$\Delta = -~\Gamma$, $\tau=400~\mu$s]: $\phi$ image. K) [$\Delta = 0~\Gamma$, $\tau=800~\mu$s]: atom column density $\rho$ ($1/\mu$m$^2$). L) [$\Delta = 0~\Gamma$, $\tau=10~$ms]: $\rho$ image during cooling. M) and N): Same as F) and G) but during cooling. \label{fig3}}
\end{figure}

\subsection{Retrieving absorption and phase shift images from holograms}

Typical holograms and reconstructed images of atom samples are displayed in Fig.~\ref{fig3}. The hologram $H$ with a small atom sample and $H_0$ in the absence of the sample are nearly identical (Fig.~\ref{fig3}A). Careful subtraction reveals the interference fringes (Fig.~\ref{fig3}B), albeit barely visible due to the photon shot noise.

As detailed in section~\ref{secDetail}, to obtain $E_r$ from $H_0=|E_r|^2$ we holographically extract $E_{r,2}$ and $\varphi_{1,2}={\rm arg}[E_{r,1}+E_{r,2}]$ , by assuming a spherical wavefront for $E_{r,1}$ from ${\bf r}_1=(x_1,y_1,0)$. With $x_1,y_1$ set as zero (validated by optical alignment to the camera center), the precise value of the ${\bf r}_1$-camera distance $L$ is calibrated by the pre-measured focal point separation $d$. Our final estimation $E_r\approx\sqrt{H_0}e^{i\varphi_{1,2}}$ partly accounts for the $E_{\rm speck}$ effect in the $E_r$ amplitude~\cite{futurework}. Any errors in the estimation of ${\bf r}_{1,2}$, as well as the spherical wave assumption itself, translate into aberration in the reconstructed image, which can become important at large NA. However, we can always minimize such errors by optimizing the reconstruction of $E_{r,2}$ at its diffraction-limited focal point ${\bf r}_2$. 

We determine the precise value of $z_0$ by optimizing the reconstruction of a density modulated atomic sample subjected to the writing lattice (Figs.~\ref{fig3}HI).  We obtain the probe field $E_{r,1}(z_0)=\hat{\rm U}_0 E_{r,1}(L)$ (Fig.~\ref{fig3}C), with $E_{\rm speck}$ ignored in this work~\cite{futurework}, to simultaneously retrieve the phase shift $\phi={\rm Im}[\log(1+E_{s}(z_0)/E_{r,1}(z_0))]$ and absorption coefficient $\alpha= |1+E_s(z_0)/E_{r,1}(z_0)|^2-1$.

\subsection{Typical phase shift, absorption, and atomic density images}

Figure~\ref{fig3}D shows a phase shift image probed at $\Delta=30~\Gamma$, where an SNR~$<1$ for the subtracted hologram on the camera plane (Fig.~\ref{fig3}B) is remarkably enhanced to SNR~$\approx 7$, due to the numerical focusing of $E_s$ when propagated from $z=L$ to $z=z_0$. The resolution $\delta\phi\approx 1.7$~mrad is a factor of 4 smaller than the minimal noise level in phase-contrast imaging using the same camera without pixel saturation~\cite{foot:Superpix}. The absorption $\alpha\approx \frac{\Gamma}{\Delta}\phi$ in Fig.~\ref{fig3}E is below the noise floor $\delta \alpha\approx 0.4\%$ and is hardly detected. Such $\delta \phi$ and $\delta \alpha$ sensitivity levels are consistently reached under various experimental conditions (Figs.~\ref{fig3}D-G,J,K,M,N) when the speckle noise induced blurring is below the shot noise level (Figs.~\ref{fig2},~\ref{fig4}), and if $N_p$ approaches $N_{\rm max}=2^{14}-1$. The opposite sign of the detunings leads to advanced (Figs.~\ref{fig2}DFHM) and retarded (Figs.~\ref{fig2}IJ) phase shift, as expected. The sensitivity can be further improved by reducing the intensity ratio $\kappa\approx 0.01$ with reduced $z_0/L$.

The atom number noise level $\delta N_{\rm atom}$ scales with $|(\frac{\Gamma}{2}+i\Delta)|/\frac{\Gamma}{2}$ with probe intensities $I$ well below the saturation intensity $I_s=36$~W/m$^2$, and is minimized at $\Delta=0$. With the absorption cross-section $\sigma=0.14~\mu$m$^2$ for the unpolarized atoms, we convert the absorption image into the sample column density $\rho$ in Fig.~\ref{fig3}K ($\Delta=0~\Gamma$, $\tau=800~\mu$s). The rms atom number noise level is $\delta N_{\rm atom}=\delta \rho A_r\approx 0.8$. To see atom shot noise, we need to fix the atoms within $A_r$ during imaging, and improve $\delta N_{\rm atom}$ further.

\subsection{Imaging during cooling}

The sample in Fig.~\ref{fig3}L is continuously probed during cooling with the probe intensity $I=0.6$~W/m$^2$ and $\tau$=10~ms . Using such a weak probe, the noise level in Fig.~\ref{fig3}L is twice the photon shot noise level due to poor subtraction of light scattered from the cooling beams.

Imaging during cooling allows us to observe the effect of light shift on the atoms by the cooling beams. For two samples of similar size, we display absorption and phase shift data for holograms taken during (Figs.~\ref{fig3}M,N) and after (Figs.~\ref{fig3}F,G) cooling. With $\Delta=1.9~\Gamma$, the absorption is relatively increased when imaged while cooling, from which we infer a light shift of $\delta\approx4$~MHz to the cooling transition so the dressed probe detuning $\tilde\Delta=\Delta-\delta$ is closer to the resonance.

Imaging during cooling is a prerequisite for number-resolving detection of weakly confined gases over long exposure times~\cite{Bakr09,Sherson10,Nelson07}. As in Figs.~\ref{fig3}M,N, we find holographic detection of atoms during laser cooling can reach photon shot-noise limited sensitivities~\cite{Velutic06}.

\subsection{Spatial resolution}

We test the resolution of the microscope by imaging atoms subjected to the 1D ``writing'' lattice. Figures~\ref{fig3}H,I are phase shift images of the density modulated sample with $\Delta=1.9~\Gamma$ and $\Delta=-7.1~\Gamma$ respectively. Here again as expected the opposite sign of the detunings leads to advanced and retarded phase shift. The $20~\mu$m fringe period is clearly resolved with the diffraction limited resolution of $x_{\rm res}=5.2~\mu$m in this work.

\subsection{Convergence of the reconstruction algorithm}
\begin{figure}[htb!]
 \centering
  \includegraphics [width=5in] {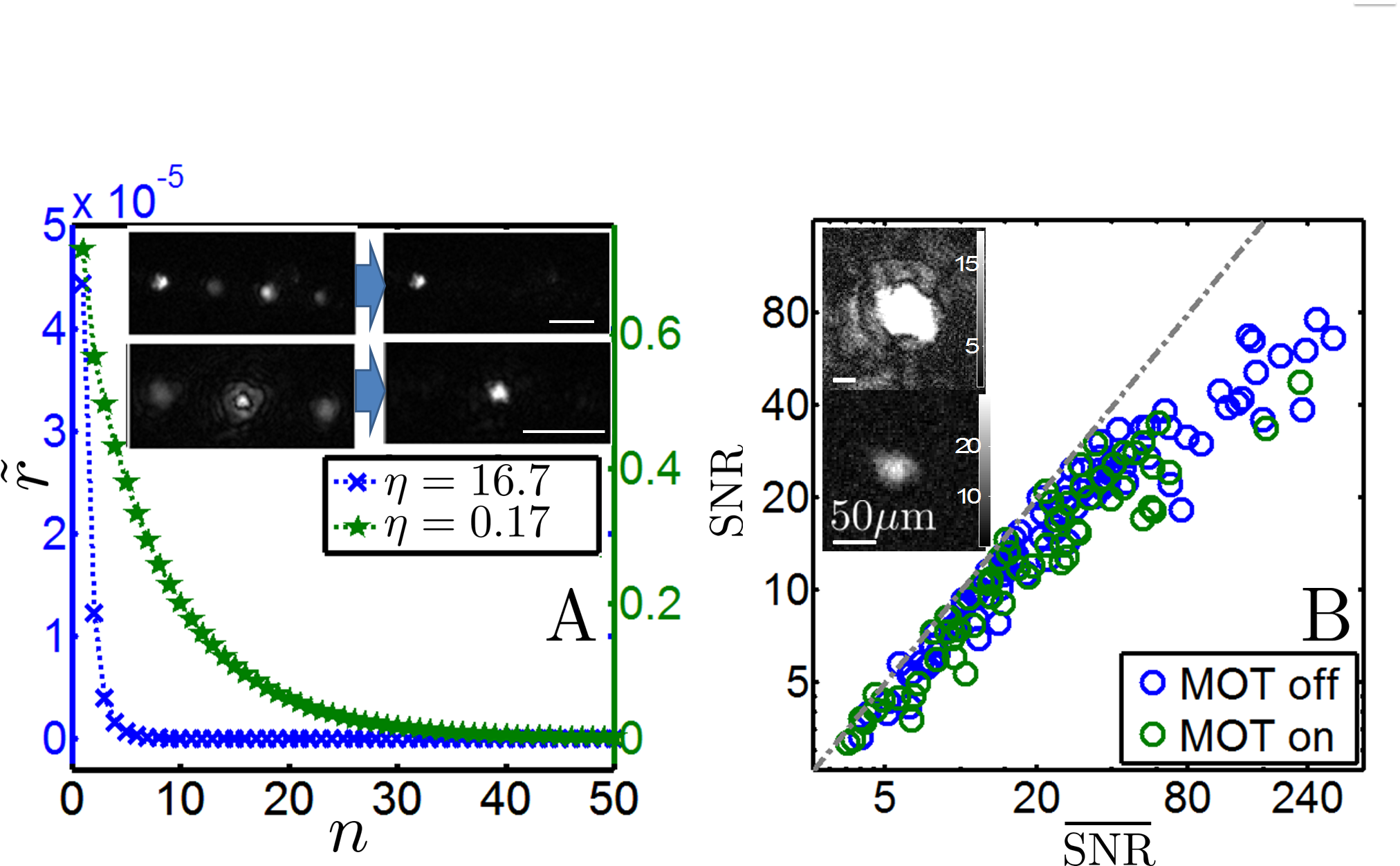}
  \caption{A) Convergence of the twin and DC image removal algorithm for the hybrid holographic microscope with different power ratios $\eta$ (${\rm F}\approx$ 3.8).  Inset): The one-step $|E_s^{(0)}|$ (left), and converged $|E_s^{(n=50)}|$ images for $\eta=16.7$ (top) and $\eta=0.17$ (bottom) (scale bar is $d=387~\mu$m). B) SNR vs $\overline{\rm SNR}$ (signal to shot noise level) for an unbiased collection of reconstructions. Inset): the converged $|E_s^{(n=50)}|$ (V/m) for a large (top) and a small (bottom) atom sample.
 \label{fig4}}
\end{figure}

The images reconstructed in Fig.~\ref{fig3} are with both twin and DC images removed. The reconstruction follows Eq.~(\ref{eq:htwo}) and the modified form of Eq.~(\ref{eq:Algorithm}) (Eq.~(\ref{eq:Algorithm2})) that converges slower, but stably removes the DC noise. The $\tilde{r}-n$ plots in Fig.~\ref{fig4}A characterize such a convergence. In contrast to the simulated results in Fig.~\ref{fig2}A where $E_s(z_0)$ is known, we calculate the residual $\tilde{r}$ defined by the difference between $E_s^{(n)}$ and $E_s^{(n=250)}$ (nearly invariant under iteration.). Comparing with the simulated results in Fig.~\ref{fig2}A, we notice that for $\eta>1$, the hybrid geometry achieves nearly full real and twin image separation ($\tilde{r}\approx 5\times 10^{-5}$) with a one-step reconstruction, a situation similar to the ``off-axis'' geometry~\cite{Leith}, but retains the convenience of measuring forward scattering at large NA in the inline geometry. Due to the modification of Eq.~(\ref{eq:Algorithm}), the convergence for $\eta=0.17$ is not as oscillatory and is twice as slow as the simulated case (Fig.~\ref{fig2}A).
\subsection{Reaching the shot noise limit}

We characterize each holographic reconstruction with two numbers, the SNR (signal to noise ratio) and $\overline {\rm SNR}$ (signal to shot noise ratio). The signal is defined as the mean value of the reconstructed $|E_s|$ above $60\%$ of its maximum. The noise level is given by $\sqrt{\langle |E_s|^2 \rangle_{\rm D}}$, where $\rm D$ is the area of an annulus surrounding $\hat {\rm P}$ with ring width equal to $a$ ($a\approx 150~\mu$m). The shot noise level in the reconstructed $E_s$ is related to the pixel counts as $\sqrt{N_p}$. As detailed in section~\ref{secDetail}, it can be approximated by $\sqrt{2\hbar \omega/A_r Q \tau}$ (i.e. 2-photon equivalent electric field amplitude) but we have performed a more precise calculation that considers the shot noise contribution from $H$ and $H_0$  separately. The $\overline {\rm SNR}$ corresponds to $\sqrt{N_s/2}$ in the simulated situation (Fig.~\ref{fig2}B). Due to the high pixel count in the holographic recording, the 8 count/pixel camera readout noise contributes negligibly to the overall noise level.

We plot SNR vs $\overline {\rm SNR}$ for an unbiased collection of holographic reconstructions in Fig.~\ref{fig4}B. With $\overline {\rm SNR}\approx \rm SNR$, our reconstructions with SNR $\lesssim 30$ frequently reach the photon shot noise limit. The few data points with SNR $\lesssim 30$ that are not photon shot-noise limited are caused by changes in the speckle pattern, and relative phase fluctuation in $E_{r,1,2}$ between the recording of $H$ and $H_0$, which are not taken into account in the subtraction (section~\ref{secDetail}). Images with larger SNR in this experiment are deteriorated by the speckle noise induced blurring, similar to the simulated situation in Fig.~\ref{fig2}B. Preliminary additional work has shown that by including $E_{\rm speck}$ into the reference field phase estimation, the speckle induced blurring can be substantially suppressed~\cite{futurework}. 

\section{Conclusion and outlook}

We have demonstrated diffraction limited holographic imaging of cold atoms with photon shot-noise limited sensitivity for the first time to our knowledge~\cite{Gross}. The phase shift and absorption sensitivities are beyond those in standard imaging if the same camera is used without binning~\cite{foot:Superpix}. The simple and robust setup is in contrast to phase-contrast imaging~\cite{phasecontrast} and spatial heterodyne imaging~\cite{Kadlecek01}, and our method can image an extended atomic sample where phase-contrast imaging is likely to be deteriorated by artifacts~\cite{artifact2}. Such detection sensitivities are also achieved during laser cooling, a prerequisite for number-resolving detection of weakly confined gases over a long exposure time\cite{Bakr09,Sherson10,Nelson07,Velutic06}. The technique can be integrated into atom chips~\cite{AtomChipScience} where nano-fabricated pinholes are back-illuminated to generate the reference fields. Increasing the information contained in $E_r$, as in the hybrid geometry, may also be useful in X-ray and electron holography~\cite{Jiang,Germann}. 

Comparing with standard absorption/phase contrast imaging, our technique has a signal that is a factor of two less. This intrinsic reduction stems from the loss of information in holography as only one quadrature of the oscillatory interference $E_r^* E_s$ is detected. In comparison, in standard imaging $E_r$ optimally interferes with $E_s$, generating the largest possible signal of either absorption or phase shift. Practically one can increase the signal amplitude by increasing the exposure time, as long as the hologram does not have a count that exceeds $N_{\rm max}$.

Although we reach the shot-noise limit for small samples, the speckle induced blurring (Figs.~\ref{fig2}B,~\ref{fig4}B) limits the magnitude of the SNR for samples with large optical depth. The effect depends on the location of the scatterers that generate the speckle noise. Our maximum SNR $\approx 30$ suggests a maximum phase shift (absorption) of 60 mrad ($12~\%$) before speckle noise noticeably affects imaging. Including $E_{\rm speck}$ into the reference field phase estimation, as in our preliminary additional work, shows substantial speckle noise suppression~\cite{futurework} for imaging samples with larger optical depth. The largest retrievable phase shift $\phi_{max }$ is related to the instability in removing the intense DC term (Eq.~(\ref{Eq:EH}), Sec.~\ref{secDC}), a topic needing further investigation. In our setup, with an extended sample where the camera is at the near field of $E_s$, we find that $\phi_{max } \approx 1.2$~rad, while for smaller samples $\phi_{max}$ can be larger.


The lengthy numerical process may be sped up with advanced algorithms~\cite{Segal10} and processing reconstructions using graphics processing units.  The atom number noise level $\delta N_{\rm atom} \propto \sqrt{x_{\rm res}^2/\sigma}$ reduces with improved imaging resolution, provided that atoms are confined within $x_{\rm res}$ during imaging. The moderate spatial resolution $x_{\rm res}=5.2~\mu$m can be improved, for example, by using a camera with a larger width $w$, or by magnifying the camera with lenses (Fig.~\ref{fig1}). The aberration in the latter case is self-suppressed due to the common optical paths for $E_s$ and $E_r$ when $z_0\ll L$, and can be further corrected numerically. With a factor of 10 increase in NA to 0.7 and using a camera with Q=0.8 (Q=0.1 in this work), we expect $\delta N_{\rm atom}\approx 0.08$ with $\tau=100~\mu$s resonant detection. By reducing $z_0$ for a condensed sample, the $N_{\rm max}$-limited sensitivity $\delta \phi$ can be improved further, even if the camera is magnified by lenses. With a spin-polarized sample to increase $\sigma$, $\delta N_{\rm atom}=0.2$ should be achievable for $\Delta=5~\Gamma$, allowing off-resonant imaging with single atom sensitivity. This could be a favorable scenario for precise phase imaging of cold atoms at high density, where the line shift/broadening due to multiple scatterings has been shown to prevent faithful and efficient resonant imaging~\cite{Chomaz12}.

We conclude by mentioning the possibility of non-destructive probing of multiple atoms in single sites of an optical lattice~\cite{Nelson07,Bakr09,Sherson10,Velutic06}. The idea is based on optical shielding~\cite{Suomien,Yurovsky}: Blue detuned light enhances the repulsion between symmetrically excited atom pairs. In the proposed scheme, atoms are trapped in single lattice sites with strong $x,y$ confinement. A blue detuned probe and auxiliary cooling beams, all with polarization in the $x-y$ plane, switch on adiabatically so that atoms confined in single sites find new equilibrium positions separated along $z$. The blue-detuned molasses should allow sub-Doppler cooling of the atom array confined in the lattice site for continuous, holographic detection of the probe light phase shift/absorption as in this work. More investigations are needed to confirm the applicability of this method, which may substantially extend the range of observables in quantum gas microscopy~\cite{Bakr09,Sherson10}.

\section{Acknowledgments}
We gratefully acknowledge early experimental contributions by M.~Worsfold and B.~Knight-Gregson, and helpful discussions with Dr. E.~Rocco, Dr.~J.~V.~Porto and Prof M.~Charlton.

\section{Appendix: Details of the holographic reconstruction\label{secDetail}}


\subsection{Extracting the reference field.\label{secReffield}} \normalfont \indent To extract the reference field $E_r$, we estimate $\varphi={\rm arg}[E_r]$ so that $E_r=\sqrt{H_0}e^{i\varphi}$.  With a large $L$ and tightly focused $E_{r,j}$ ($j=1,2$), we assume the wavefronts $E_{r,j}({\bf R})\propto\sqrt{H_{j}}e^{i k |{\bf R}-{\bf r}_j|}$ with ${\bf R}=(x,y,L)$ on the camera plane to be well-approximated by spherical waves from the focal points ${\bf r}_j$. Here $H_{1}$ is an intensity image taken when only the inline source $E_{r,1}$ is on, and similarly for $H_2$ when only the off-axis source $E_{r,2}$ is on. In the following we describe the procedure to determine the location of the focal points ${\bf r}_1$ and ${\bf r}_2$, and additional adjustments to estimate $E_r(L)$ for image reconstruction, as well as to estimate $E_r(z_0)$ for phase shift $\phi$ and absorption coefficient $\alpha$ retrieval.


We first adjust the camera position so that the $H_1=|E_{r,1}|^2$ intensity is centered. We then fit $H_1$ with a 2D Gaussian $H_{1,f}$ and set its center to $x=y=0$. With an estimate of $L$, the location ${\bf r}_2$ is retrieved by propagating $E=H_0/E_{r,1}^*$ to its focal plane (with prior knowledge of the ${\bf r}_{1,2}$ relative position, the twin image at $-{\bf r}_2$ is easily identified). The displacement $|{\bf r}_2-{\bf r}_1|=d=387.0~\mu$m, measured before the installation, is used to calibrate the $L$ parameter in $E_{r,1}$.


Reconstruction of the $E_{r,2}$ focal point also provides an estimate of the relative phase $\phi_r$ between $E_{r,2}$ and $E_{r,1}$. We therefore reach an estimate $E_r^{(0)}(a_1,a_2,\phi)=a_1 \sqrt{H_{1,f}}e^{i k |{\bf R}- {\bf r}_1|}+a_2 \sqrt{H_{2,f}}e^{i k |{\bf R}-{\bf r}_2|+i\phi_r}$ at the camera plane with $a_1=a_2=1$ ($H_{2,f}$ is a 2D fit of the $H_2$ intensity.). To improve the accuracy of the estimation, we minimize the difference $H_0-|E_r^{(0)}(a_1,a_2,\phi_r)|^2$ to achieve the optimal $E_{r,{\rm opt}}^{(0)}=a_{1,\rm opt} \sqrt{H_{1,f}}e^{i k |{\bf R}- {\bf r}_1|}+a_{2,\rm opt} \sqrt{H_{2,f}}e^{i k |{\bf R}-{\bf r}_2|+i\phi_{r,\rm opt}}$. Finally, we use $\varphi_{1,2}={\rm arg}[E_{r,{\rm opt}}^{(0)}]$ to approximate $\varphi={\rm arg}[E_r]$ in this work so that $E_r(L)\approx \sqrt{H_0}e^{i\varphi_{1,2}}$.

From the estimated $E_r(L) $ it is straightforward to calculate $E_r(z_0)=\hat {\rm U}_0 E_r(L)$. To avoid FFT boundary artifacts, we expand the fitted $H_{1,f}$ and $H_{2,f}$ onto a grid twice as large as the camera, so both fitted intensities are well-contained within the grid. Similar to the $E_r(L)$ estimation, the approximation $E_r(z_0) \approx \hat {\rm U}_0 E_{r,{\rm opt}}^{(0)}$ ignores speckle noise.


\subsection{Iterative twin and DC removal.\label{secDC}} \normalfont \indent Eq.~(2) assumes $|E_s|^2\ll |E_r|^2$ over the camera. To remove the DC term Eq. (2) is modified, using our estimation of $E_s$ at each iteration $E_s^{(n)}$, we calculate the DC term $\approx |{\rm\hat{U}_0^{-1}} {\rm\hat{P}} E_s^{(n)}|^2$ and subtract it from the hologram.
\begin{equation}
\begin{array}{c}
E_{s}^{(0)} = \hat {\rm U}_0 E_{H},\\
\tilde{E}_{s}^{(0,n)} = \hat{\rm{U}}_0 \left( E_H  - | \hat{\rm{U}}_0^{-1} \hat{\rm{P}} E_s^{(n)} |^2/E_r^* \right),\\
E_{s}^{(n+1)} =  (1-\nu) \hat {\rm P}{E}_s^{(n)} + \nu \hat {\rm U}_0 \hat {\rm C} \hat {\rm U}_0^{-1} \left( \tilde{E}_{s}^{(0,n)} -
\hat {\rm P} {E}_{s}^{(n)} \right).
\end{array}
\label{eq:Algorithm2}
\end{equation}

We introduce $\nu$ to control the speed, in Eq.~(2) $\nu=1$. The DC removal makes the algorithm nonlinear, and we find it can be unstable when $E_s$ is large. Adjusting $\nu$ improves the stability and we set $\nu=0.5$ in this work. The convergence is a factor of two slower (Fig. 4A), while the oscillatory feature in (Fig. 2A) is suppressed.

\subsection{Optimal background subtraction.} \normalfont\indent  
$H$, $H_0$, $H_1$ and $H_2$, are affected by laser power and phase fluctuations, and ambient light.
For retrieving $E_r$ using $H_0$, $H_1$ and $H_2$ as in section~\ref{secReffield}. We take a background image $B_0$ with the same exposure time as $H_0$ and subtract it directly. The subtraction $H-H_0$ in Eq. (1), requires greater precision. We have developed a two-step optimization for the subtraction as following.

First, using the two point interference, we optimize the subtraction $H_R^{(0)}(b)=H- b H_0$, with a parameter $b$ to account for power fluctuation between recordings. The optimization starts with the complex field $E(b)=H_R^{(0)}(b)/E_{r,1}^{*}$ which is propagated to $z=0$, $\hat {\rm U}(-L)E(b)$, where three spots are formed. We isolate the one at ${\bf r_2}=(d, 0, 0)$ and calculate its residual power $dP(b)$ (The others correspond to the inline and off-axis twin sources.). $dP(b)$ is minimized at $b=b_{\rm opt}$. To eliminate atomic signal from affecting $b_{\rm opt}$, we apply a mask that blocks the geometric shadow of the sample in $H$ and $H_0$.


\indent Optimization based on the fringes cannot account for a relative $E_{r,1}$, $E_{r,2}$ power fluctuation. In addition, atomic fluorescence causes a uniform background $f_{\rm bk}$. We account for these, together with the ambient light, using a four parameter ($c_j$, $j=1,2,3,4$) optimization.
\begin{equation}
H_R=H - b_{\rm opt} H_0 - c_1 B  + c_2 B_0 - c_3 f_{\rm bk} - c_4 P_f.
\label{eq:Optimization}
\end{equation}
$P_f=H_{f,1}-H_{f,2}$ accounts for a relative power fluctuation and $B$ is the ambient background for $H$.

We then propagate $E(c_j) = H_R(c_j)/E_r^*$ via $\hat {\rm U}_0$ to $z=z_0$ where we expect the sample's in-focus image $E_s(z_0)$ and its out-of-focus twin. We use $1-\hat {\rm P}$ to exclude $E_s$. The leftover twin image is converted into a real image and refocused to $z=z_0$ using the $\hat {\rm U}_0 \hat {\rm C} \hat {\rm U}_0^{-1}$ operation as in Eq.~(2), before it is also excluded. This ensures the final image is nearly free of atomic signal, from which we calculate the rms noise level $N(c_j)$ to be minimized at $c_{j,{\rm opt}}$ and thus the final $H_{R,{\rm opt}}$ which is ``$H-H_0$'' in Eq.~(1). This takes $\approx 10$ minutes with a PC (Intel Core i5-2400 CPU, 3.1 GHz).

\subsection{Propagation of shot noise.} \normalfont\indent 
The rms shot noise level in the holograms, proportional to the square root of the pixel count $N_p=H A_p Q \tau/\hbar \omega$, is given by $n_H=H/\sqrt{H A_p Q \tau/\hbar \omega}$ and similarly $n_{H0}=H_0/\sqrt{H_0 A_p Q \tau/\hbar \omega}$. The associated shot noise $\delta E_H$ of $E_H$ in Eq.~(1), in each pixel, has a rms level:

\begin{equation}
\begin{array}{c}
n_{EH}=|\sqrt{n_H^2+n_{H0}^2}/E_{r}^*|,\\
=\sqrt{2\hbar \omega/A_p Q \tau}.
\end{array}
\label{eq:noise1}
\end{equation}
Here we assume $n_H=n_{H0}$. To derive the rms level of the shot noise field $\delta E_a$ at $z=z_0$, we evaluate the intensity contribution $|\hat {\rm U}_0 \delta E_H|^2$ from each pixel on which $\delta E_H$ has a random but uniform amplitude. The contribution is summed over the pixels to give the intensity of $\delta E_a$.

More specifically, we consider the field $\delta E_H$ through each pixel, with wavefront area restricted by $A_p$, as spherical waves propagating toward ${\bf r}_1$ and ${\bf r}_2$. We calculate the noise level of $E_s$ near the sources, ${\bf r}=O(d)$, via Kirchhoff's diffraction theorem and further a 2D integration of the intensity contribution over the pixels.
\begin{equation}
\begin{array}{c}
n_{E,{\bf r}=O(d)}=\sqrt{2\hbar \omega/\tilde A_r Q \tau},\\
\tilde A_r=\lambda^2/(4 {\rm NA\times Tan^{-1}(NA)}).
\end{array}
\label{eq:noise2}
\end{equation}
Here ${\rm NA}=w/\sqrt{w^2 + 4 L^2}$, with $z_0\ll L$ the noise level $n_{E,a}\approx n_{E,{\bf r}=O(d)}$. With $\tilde A_r \approx A_r$, The rms level of shot noise $\delta E_a$ near the atom location is

\begin{equation}
n_{E,a}\approx \sqrt{2\hbar \omega/A_r Q \tau}.
\label{eq:noise3}
\end{equation}

\subsection{The final shot noise level after iteration}

Shot noise modifies the input of Eq.~(2) to $\tilde E_s^{(0)}=E_s^{(0)}+\delta E_a$, while $E_H$ is modified to $\tilde E_H=E_H+\delta E_H$. As the iteration converges to $\tilde E_s=E_s(z_0)+\delta E_s$, considering $|E_s|^2 \ll |E_r|^2$, it is easy to show the following:
\begin{equation}
\delta E_s=\delta E_a-\hat {\rm U}_0 \hat {\rm C} \hat {\rm U}_0^{-1}\hat {\rm P} \delta E_s.
\label{eq:noise4}
\end{equation}

The final noise $\delta E_s$, as a functional of the shot noise $\delta E_a$ specified by Eq.~(\ref{eq:noise3}),
is difficult to calculate analytically. However, if $\hat {\rm P} \delta E_s$ is, as verified numerically, a uniform shot noise pattern filtered by the aperture $\hat {\rm P}$, while the twin image $\hat {\rm U}_0 \hat {\rm C} \hat {\rm U}_0^{-1}{\rm P} \delta E_s$ is not correlated with the shot noise $\delta E_a$ in Eq.~(\ref{eq:noise4}), then:
\begin{equation}
n_{Es}^2=n_{E,a}^2+\xi^2 n_{Es}^2.
\label{eq:noise5}
\end{equation}
Here $\xi$ is the ratio of the rms level between $\hat {\rm U}_0 \hat {\rm C} \hat {\rm U}_0^{-1}\hat {\rm P} \delta E_s$ and  $\delta E_s$. And the final shot noise level becomes:
\begin{equation}
n_{Es}\approx \frac{\sqrt{2\hbar \omega/A_r Q \tau}}{\sqrt{1-\xi^2}}.
\label{eq:noise6}
\end{equation}
For the inline geometry $\xi_{\rm inline}=1/(1+\frac{2w/L}{a/z_0})$ for ($z_0\ll L$), which is determined by the area ratio of the apertured shot noise given by $\hat{{\rm P}}$ and its NA-limited shadow at the twin image plane. For the hybrid geometry $\xi\approx (1-\eta) \xi_{\rm inline}$ for $\eta<1$. In this work $\xi \sim 0.1$, the increase of the shot noise from $n_{Ea}$ to $n_{Es}$ due to the algorithm is $< 1\%$.

\subsection{Shot-noise limited {\rm SNR}, and $N_{\rm max}$-limited sensitivity}

With the noise level of $E_s$ specified by Eq.~(\ref{eq:noise6}), the ${\rm SNR}= \sqrt{\langle |E_s(z_0)/n_{Es}|^2 \rangle}$ over a particular $A_r$, is ${\rm SNR}=\sqrt{N_s/2}$ with $N_s= \langle |E_s(z_0)|^2 \rangle A_r Q\tau/\hbar \omega$ (we consider $\xi^2/2 \ll 1$.).

But, for a camera with $N_{\rm max}$, the relation $N_p=|E_r(L)|^2  A_p Q\tau/\hbar \omega<N_{\rm max}$ must be satisfied. By considering the ${\rm SNR}=1$ limit, the minimum detectable $E_s$ must obey $\left |\frac{E_s(z_0)}{E_r(L)}\right |>\sqrt{\frac{2 A_p}{N_{\rm max} A_r}}$, or, with $\kappa=|E_r(L)|^2/|E_r(z_0)|^2$,

\begin{equation}
 |E_s(z_0)|>\sqrt{\frac{2 \kappa A_p }{N_{\rm max} A_r}}|E_r(z_0)|.
\label{eq:sensitivity}
\end{equation}
From Eq.~(\ref{eq:sensitivity}) we obtain the phase shift $\delta \phi=\sqrt{\frac{2 \kappa A_p }{N_{\rm max} A_r}}$ and absorption coefficient $\delta \alpha = 2\sqrt{\frac{2 \kappa A_p }{N_{\rm max} A_r}}$ sensitivities.


\section*{References}
\begin{thebibliography}{10}

\bibitem{BECpub}
M.~H.~Anderson, J.~R.~Ensher, M.~R.~Matthews, C.~E.~Wieman and E.~A.~Cornell, Science {\bf 269}, 198, 1995;
K.~B.~Davis, M.~O.~Mewes, M.~R.~Andrews, N.~J.~van Druten, D.~S.~Durfee, D.~M.~Kurn and W.~Ketterle, Phys. Rev. Lett. {\bf 75}, 3969
1995;
C.~C.~Bradley, C.~A.~Sackett, J.~J.~Tollett, and R.~G.~Hulet, Phys. Rev. Lett.
{\bf 75}, 1687, 1995.

\bibitem{ImageBreakthrough}
J. M. Gerton, D. Strekalov, I. Prodan and R. G. Hulet Nature {\bf 408}, 692695, 2000;
L.~E.~Sadler, J.~M.~Higbie, S.~R.~Leslie, M.~Vengalattore and D.~M.~Stamper-Kurn,
Nature {\bf 443}, 312, 2006; N.~Gemelke, X.~B.~Zhang, C.~L.~Hung and C.~Chin, Nature
{\bf 460}, 995-998, 2009.



\bibitem{Bakr09}
W.~S.~Bakr, J.~I.~Gillen, A.~Peng, S.~F\"olling and M.~A.~Greiner, Nature {\bf 462},
74, 2009.

\bibitem{Sherson10}
J.~F.~Sherson, C.~Weitenberg, M.~Endres, M.~Cheneau, I.~Bloch and S.~Kuhr, Nature
{\bf 467},68, 2010.

\bibitem{Nelson07} K.~D.~Nelson, X.~Li and D.~S.~Weiss, Nature Phys. {\bf 3}, 556,
2007.

\bibitem{Chomaz12}
L.~Chomaz, L.~Corman, T.~Yefsah, R.~Desbuquois, and J.~Dalibard, New J. Phys. {\bf 14},
055001, 2012.

\bibitem{Yefsah11} T.~Yefsah, R.~Desbuquois, L.~Chomaz, J.~K.~G\"unter,
and J.~Dalibard, Phys. Rev. Lett. {\bf 107}, 130401, 2011.



\bibitem{Othersolution} Other solutions include fitting the tails of the image or to repeatedly image a fraction of the atomic sample, e. g., E.~R.~S.~Guajardo, M.~K.~Tey, L.~A.~Sidorenkov, and R.~Grimm, Phys. Rev. A {\bf 87}, 063601, 2013; A.~Ramanathan, S.~R.~Muniz, K.~C.~Wright, R.~P.~Anderson, W.~D.~Phillips, K.~Helmerson and G.~K.~Campbell, Rev. Sci. Instrument {\bf 83},
083119, 2012.

\bibitem{phasecontrast}
M.~R.~Andrews, D.~M.~Kurn, H.-J.~Miesner, D.~S.~Durfee, C.~G.~Townsend, S.~Inouye,
and W.~Ketterle, Phys. Rev. Lett. {\bf 79}, 553, 1997.

\bibitem{Higbie05}
J. M. Higbie, L. E. Sadler, S. Inouye, A. P. Chikkatur, S. R. Leslie, K. L. Moore,
V. Savalli, and D. M. Stamper-Kurn, Phys. Rev. Lett. {\bf 95}, 050401, 2005.

\bibitem{Sanner11}
C.~Sanner, E.~J.~Su, A.~Keshet, W.~Huang, J.~Gillen, R.~Gommers, and W.~Ketterle, Phys. Rev. Lett. {\bf 106}, 010402, 2011.

\bibitem{Schley13}
R.~Schley, A.~Berkovitz, S.~Rinott, I.~Shammass, A.~Blumkin, and J.~Steinhauer,
Phys. Rev. Lett. {\bf 111}, 055301, 2013.

\bibitem{Kadlecek01} 
S.~Kadlecek, J.~Sebby, R.~Newell, and T.~G.~Walker, Opt. Lett. {\bf 26}, 137, 2001.

\bibitem{Turner04}
L.~D.~Turner, K.~P.~Weber, D.~Paganin, and R.~E.~Scholten, Opt. Lett. {\bf 29}, 232, 2004.


\bibitem{Turner}
L.~D.~Turner, K.~F.~E.~M.~Domen, and R.~E.~Scholten, Phys. Rev. A {\bf 72}, 031403, 2005.

\bibitem{TurnerThesis}
L. D. Turner, ``Holographic Imaging of Cold Atoms'', Ph.D. dissertation, University of Melbourne, Australia, 2004.

\bibitem{Light}
P.~S.~Light, C.~Perrella, and A.~N.~Luiten, Appl. Phys. Lett. {\bf 102}, 171108, 2013.

\bibitem{Gabor48} D.~Gabor, Nature {\bf 161}, 777, 1948.

\bibitem{Kanka11} M.~Kanka R. Riesenberg, P. Petruck and C. Graulig,  Opt. Lett., {\bf 36}, 3651, 2011.

\bibitem{Gross}
M.~Gross and M.~Atlan, 
Opt. Lett., {\bf 32}, 8, 2007. 

\bibitem{Huang14}
C.~Y.~Huang, H.~S.~Chen, C.~Y.~Liu, C.~H.~Chen and D.~J.~Han, J. Opt. Soc. Am. B, {\bf 31}, 1, 87, 2014.

\bibitem{Ku11}
T.~P.~Ku, C.~Y.~Huang, B.~W.~Shiau and D.~J.~Han, Opt. Express, {\bf 19}, 4, 3730, 2011.

\bibitem{Yamaguchi97}
I.~Yamaguchi and T.~Zhang, Opt. Lett. {\bf 22}, 16, 1997; 


\bibitem{foot:Superpix} In both standard and holographic imaging, $N_{\rm max}$ can be effectively increased through pixel and frame binning.


\bibitem{review1}
J. R. Fienup, Appl. Opt. {\bf 21}, 2758, 1982. 

\bibitem{koren1991}
G.~Koren, D.~Joyeux and F.~Polack, Opt. Lett. {\bf 16}, 1979, 1991.


\bibitem{Latychevskaia}
T. Latychevskaia and H. W. Fink, Phys. Rev. Lett. {\bf 98}, 23, 233901, 2007.

\bibitem{Rong}
L. Rong, Y. Li, S. Liu, W. Xiao, F. Pan, D. Wang, Opt. Laser. Eng, {\bf 51}, 5, 553, 2013.  


\bibitem{foot:notation}
To be concise we may represent $E(x,y,z)$ with $E(z)$.

\bibitem{Goodman}
{\it Introduction to Fourier Optics Third Edition}, J. W. Goodman (Roberts and Company, 2005).

\bibitem{lye1}
J.~J.~Hope and J.~D.~Close, Phys. Rev. Lett. {\bf 93}, 180402, 2004; Phys. Rev. A {\bf 71}, 043822, 2005.




\bibitem{futurework}
To include $E_{\rm speck}$ in the estimation of $\varphi$, one can locate the distant point-like scatterers holographically. We have developed an algorithm for such inclusion that shows suppression of the speckle-induced blurring as in Figs.~\ref{fig2}B,~\ref{fig4}B. The results will be published elsewhere.

\bibitem{metcalf}
{\it Laser Cooling and Trapping}, H.~Metcalf and P.~van~der~Straten (Springer-Verlag, 1999).




\bibitem{Velutic06}
When detecting multiple atoms, holographic imaging shares the advantage of collective enhancement of forward scattering with absorption imaging. See {\it e.g.}, I.~Tepor {\it et al}, Phys. Rev. Lett. {\bf 97}, 023002, 2006.

\bibitem{Leith}
E.~N.~Leith and J.~Upatnieks, J.Opt. Soc. Am. {\bf 54}, 1295, 1964.


\bibitem{artifact2}
A.~H.~Bennett, H.~Osterberg, H.~Jupnik, and O.~W.~Richards, {\it Phase microscopy; principles and applications} (Wiley \& Sons, New York USA, 1951).

\bibitem{AtomChipScience} J.~Fortagh and C.~Zimmermann, Science {\bf 307}, 860, 2005.

\bibitem{Jiang}
H.~Jiang et al, Phys. Rev. Lett. {\bf 110}, 205501, 2013.

\bibitem{Germann}
M.~Germann, T.~Latychevskaia, C.~Escher, and H.~Fink, Phys. Rev. Lett. {\bf 104}, 095501, 2010.

\bibitem{Segal10}
S.~R.~Segal,  Q.~Diot, E.~A.~Cornell, A.~A.~Zozulya, and D.~Z.~Anderson, Phys. Rev. A {\bf 81}, 053601, 2010.





\bibitem{Suomien}
K.~Suominen, M.~Holland, K.~Burnett and P.~Julienne, 
Phys. Rev. A {\bf 51}, 1446, 1995;
\bibitem{Yurovsky}
V.~A.~Yurovsky and A.~Ben-Reuven, 
Phys. Rev. A {\bf 55}, 3772, 1997.




\end{thebibliography}
\end{document}